\documentclass[prd,amsmath,amssymb,aps,longbibliography,superscriptaddress,twocolumn,nofootinbib,10pt,showkeys]{revtex4-1}
\usepackage{textcase}
\usepackage{amssymb,amsmath,mathtools} 
\usepackage{url}
\usepackage{natbib}

\usepackage{dcolumn}
\usepackage{bm}
\usepackage[T1]{fontenc}
\usepackage[utf8]{inputenc}
\usepackage{booktabs}
\usepackage{multirow}
\usepackage{graphicx}
\usepackage{xcolor}
\usepackage{microtype} 
\usepackage{hyperref}

\hypersetup{
  colorlinks=true,
  linkcolor=red,
  citecolor=blue,
  urlcolor=blue
}

\begin{document}

\title{Testing General Relativity on Galactic Scales via DESI-BAO and Strong Lensing: Circumventing Assumptions on the Hubble Constant, Sound Horizon, and Dark Energy}

\author{Hengyu Wu}
\affiliation{School of Physics and Optoelectronic, Yangtze University, Jingzhou 434023, China;}
\author{Tonghua Liu}
\email{liutongh@yangtzeu.edu.cn}
\affiliation{School of Physics and Optoelectronic, Yangtze University, Jingzhou 434023, China;}
\author{Chenggang Shao}
\email{cgshao@hust.edu.cn}
\affiliation{School of Physics and Optoelectronic, Yangtze University, Jingzhou 434023, China;}

\begin{abstract}
We present a cosmological model-independent framework for testing general relativity (GR) on galactic scales by combining baryon acoustic oscillation (BAO) angular scale measurements with 120 galaxy-scale strong gravitational lensing systems. Using artificial neural networks (ANNs) and cubic spline reconstruction, we reconstruct the BAO angular scale from SDSS, BOSS, eBOSS, and DESI Data Release 2 (DR2), and infer the angular diameter distances to lenses and sources. Crucially, All the quantities used in the GR test are derived from observations and are independent of  cosmological parameters such as the Hubble constant, the sound horizon, or the dark energy equation of state, minimizing potential biases from model-dependent distance priors. These distances are then incorporated into the strong lensing likelihood to constrain the parameterized post-Newtonian (PPN) parameter $\gamma_{\rm PPN}$ under two lens mass models: a constant-density-slope model ($P_1$) and a redshift-evolving model ($P_2$).
For the $P_1$ model, the ANN reconstruction yields $\gamma_{\rm PPN} = 1.102^{+0.148}_{-0.125}$, consistent with GR at $1\sigma$ confidence level, while the cubic spline gives $\gamma_{\rm PPN} = 1.150^{+0.139}_{-0.118}$, consistent with GR at $2\sigma$ confidence level. For the $P_2$ model, the ANN reconstruction gives $\gamma_{\rm PPN} = 1.315^{+0.181}_{-0.155}$, compatible with GR at $2\sigma$, while the spline gives $\gamma_{\rm PPN} = 1.485^{+0.193}_{-0.168}$, showing mild tension at $\sim2.5\sigma$. The constraints exhibit a clear dependence on the adopted lens mass model, underscoring the critical role of lens modeling. No significant correlation is observed between $\gamma_{\rm PPN}$ and the Einstein radius. Overall, current galaxy-scale observations are consistent with GR, providing no evidence for deviations from Einstein's theory on kiloparsec scales.
\end{abstract}

\maketitle

\section{Introduction}

Understanding the expansion history of the Universe and testing the law of gravity on cosmological scales are among the central goals of modern cosmology. Einstein's theory of GR has served as the standard description of gravity for more than a century and has passed numerous experimental tests with remarkable precision. Classical experiments such as the perihelion precession of Mercury, the deflection of light by the Sun, and the Shapiro time delay all show excellent agreement with the predictions of GR \cite{1920RSPTA.220..291D,1960PhRvL...4..337P,1964PhRvL..13..789S,1979Natur.277..437T}. With the rapid development of modern astronomical observations, these tests have reached extremely high precision within the solar system. Nevertheless, the validity of GR on galactic and cosmological scales remains less well constrained. Despite such stringent constraints in the solar system, current limits on $\gamma_{\rm PPN}$ from large-scale observations such as weak gravitational lensing, redshift-space distortions, and galaxy cluster mass profiles typically reach only the $\sim20\%$ level \cite{2011PhRvD..84h3523S,2013MNRAS.429.2249S,2016MNRAS.456.2806B,2015MNRAS.452.1171W,2016JCAP...04..023P}, which is far less precise than solar-system tests. Improving these constraints is therefore essential for verifying the validity of GR on galactic and cosmological scales. Moreover, this issue is closely related to one of the central problems in modern cosmology, namely the origin of the accelerated expansion of the Universe.

Since the discovery of cosmic acceleration from Type Ia supernova observations, a variety of theoretical explanations have been proposed. The simplest explanation is to introduce dark energy within the framework of GR, such as the cosmological constant $\Lambda$ or dynamical dark energy models like quintessence scalar fields. However, most phenomenological models of dark energy suffer, to varying degrees, from theoretical difficulties \cite{1989RvMP...61....1W,2001LRR.....4....1C,2006IJMPD..15.1753C}. These anomalies have motivated a wealth of modified gravity theories as alternatives to GR, including $f(R)$ gravity, the Dvali--Gabadadze--Porrati (DGP) braneworld model, and scalar--tensor gravity theories \cite{2010RvMP...82..451S,2016RAA....16...22Q,2017EPJC...77..502Q,2021ApJ...911..135P,2009PhRvD..79l4019B,2014PhLB..739..102Z,2015PhLB..744..213L,2015FrPhy..10j9501G,2018PhLB..779..473Z}. These theories generally predict gravitational behaviors different from GR on cosmological scales and may lead to deviations of the PPN parameter $\gamma_{\rm PPN}$ from its GR value. Therefore, precise measurements of $\gamma_{\rm PPN}$ on extragalactic and cosmological scales not only provide an important test of GR, but also offer a potential way to distinguish between dark energy scenarios and modified gravity models.

Strong gravitational lensing (SGL) is a key prediction of GR \cite{1936Sci....84..506E,2024SSRv..220...12S} and has developed into an invaluable tool in modern astronomy. SGL systems provide a powerful astrophysical tool for testing gravity on kiloparsec scales \cite{2017JCAP...07..045C,2024PhRvD.109h4074L}. When light from a background source passes near a foreground galaxy acting as a gravitational lens, multiple images, arcs, or even Einstein rings may form with angular separations close to the Einstein radius. Theoretically, the Einstein radius depends not only on the mass distribution of the lens galaxy but also on the PPN parameter $\gamma_{\rm PPN}$ and the angular diameter distances between the observer, lens, and source \cite{2010ARA&A..48...87T}. By combining measurements of the Einstein radius with stellar velocity dispersion observations of the lens galaxy, one can directly constrain $\gamma_{\rm PPN}$ beyond the solar system. Early studies using the Sloan Lens ACS Survey (SLACS) sample provided initial constraints on $\gamma_{\rm PPN}$ \cite{2006ApJ...638..703B,2006PhRvD..74f1501B}. With the increasing number of well-observed lensing systems, SGL has become an increasingly important probe for testing gravity theories on galactic scales \cite{2017ApJ...835...92C,2018Sci...360.1342C,2024MNRAS.528.1354L,2019ApJ...880...50Y}.

On the other hand, BAO provide a robust standard ruler embedded in the large-scale structure of the Universe. BAO originate from sound waves in the photon--baryon plasma of the early Universe and become frozen at the drag epoch, leaving a characteristic scale known as the sound horizon in the matter distribution \cite{2005ApJ...633..560E,2005MNRAS.362..505C}. This scale allows measurements of cosmological distances such as the angular diameter distance and the Hubble parameter. Since the first detections of BAO signals by the Sloan Digital Sky Survey (SDSS) and the 2dF Galaxy Redshift Survey, a series of large spectroscopic surveys including 6dFGS, BOSS, eBOSS, and the recent DESI survey have significantly improved the precision of BAO measurements across a wide redshift range \cite{2011MNRAS.416.3017B,2017MNRAS.470.2617A,2021PhRvD.103h3533A,2012MNRAS.425..405B}. These observations provide valuable information for constraining cosmological parameters and investigating the expansion history of the Universe.

In recent years, considerable efforts have been devoted to developing model-independent approaches in cosmological analyses. Many traditional methods rely on a specific cosmological model, such as the $\Lambda$CDM model, to provide distance information, which may introduce potential biases when testing gravity theories. To alleviate this issue, several non-parametric reconstruction techniques have been proposed, including Gaussian processes and Cubic Spline (CS) reconstruction, which allow cosmological functions to be reconstructed directly from observational data without assuming a specific background model \cite{2022ApJ...927...28L,2022ApJ...927L...1W,2025ApJ...981L..24L}. Based on these reconstructed distances, some studies have combined them with SGL observations to constrain the PPN parameter $\gamma_{\rm PPN}$ or the cosmic curvature \cite{2017ApJ...835...92C}. In addition, SGL systems with time-delay measurements have been widely used to infer cosmological distances and calibrate cosmological probes \cite{2020A&A...642A.193M,2025PhRvD.111b3524L,2020A&A...639A.101M}.

Motivated by these developments, in this work we propose a model-independent method to test GR by combining BAO and SGL observations. We first reconstruct the evolution of the BAO angular scale as a function of redshift using two non-parametric techniques: ANN and CS reconstruction. These methods allow us to obtain angular diameter distance information directly from observational data without assuming any cosmological model. We then combine the reconstructed BAO distances with measurements of Einstein radii and stellar velocity dispersions from SGL systems. By constructing the corresponding likelihood function, we obtain constraints on the PPN parameter $\gamma_{\rm PPN}$ and test the validity of GR on cosmological scales. Compared with traditional parameterized approaches, our method reduces model dependence and potential systematic biases, providing a more robust test of gravity.

This paper is structured as follows. Section II describes the observational data and methodology used in our analysis, including the gravitational-dynamical mass combination, the strong lensing and BAO datasets, the reconstruction techniques for the BAO angular scale, and the construction of the likelihood function. Section III discusses the main results and their implications. Section IV summarizes our conclusions.

\section{Data and Methodology}
\subsection{Gravitational-Dynamical Mass Combination Method}
Within the Einstein radius, the gravitational mass equals the dynamical mass. For a lens system in dynamical equilibrium, the total projected mass inferred from SGL effects should be consistent with the dynamical mass derived from the galaxy velocity dispersion within the characteristic scale of the Einstein radius \cite{2019MNRAS.488.3745C}:
\begin{equation} \label{eq:Mgrl_Mdyn}
M_{\rm{grl}}^E = M_{\rm{dyn}}^E,
\end{equation}
where $M_{\rm{grl}}^E$ and $M_{\rm{dyn}}^E$ denote the gravitational and dynamical mass, respectively, whose consistency is guaranteed by GR. In the weak-field regime, the Schwarzschild metric for a point mass $M$ can be expressed as:
\begin{equation} 
\mathrm{d}s^2 = c^2 \mathrm{d}t^2 \left(1 - \frac{2GM}{c^2r}\right) - \mathrm{d}r^2 \left(1 + \frac{2\gamma_{\text{PPN}}GM}{c^2r}\right) - r^2 \mathrm{d}\Omega^2,
\end{equation}
where $\gamma_{\text{PPN}}$ is the PPN parameter, and $\Omega$ denotes the invariant plane orbital angle, with $\gamma_{\text{PPN}}$ taking the theoretical value of unity in GR. In gravitational lensing, the gravitational mass $M_{\text{grl}}^E$ is related to the Einstein angle $\theta_E$ through a well-defined relation. Combining these two relations, one can derive the connection between the PPN parameter $\gamma_{\text{PPN}}$ and the Einstein angle $\theta_E$:
\begin{equation} \label{eq:3}
\theta_E = \sqrt{\frac{1 + \gamma_{\text{PPN}}}{2}} \left( \frac{4GM_E^{\text{grl}}}{c^2} \frac{D_{ls}}{D_s D_l} \right)^{1/2},
\end{equation}
here, $D_s$, $D_l$, and $D_{ls}$ denote the angular diameter distances between observer and source, observer and lens, and lens and source, respectively. Introducing the physical Einstein radius $R_E = D_l \theta_E$ perpendicular to the line of sight, and noting that the gravitational mass $M_E$ within the Einstein radius satisfies $M_E = M_{\rm grl}^E = M_{\rm dyn}^E$, we combine Eq.~\eqref{eq:3} with the relation $\theta_E = R_E/D_l$ to obtain:
\begin{equation}
\frac{GM_E}{R_E} = \frac{2}{1+\gamma_{\rm PPN}} \cdot \frac{c^2}{4} \cdot \frac{D_s}{D_{ls}} \theta_E.
\label{eq:M_R_relation}
\end{equation}

The dynamical mass $M_{\text{dyn}}^E$ of the lens galaxy is inferred from spectroscopic measurements of stellar velocity dispersion, under an assumed mass distribution model. Its determination depends on cosmological distances as well as the parameters characterizing the mass profile. Here we employ the general power-law density model proposed by Koopmans for $E/S0$ lens galaxies \cite{2006EAS....20..161K}:
\begin{equation}\label{eq:lens_mass_model}
\begin{cases}
\rho(r) =  \rho_0 \left( \frac{r}{r_0} \right)^{-\gamma}, \\
\nu(r) = \nu_0 \left( \frac{r}{r_0} \right)^{-\delta}, \\
\beta(r) =  1 - \frac{\sigma_\theta^2}{\sigma_r^2},
\end{cases}
\end{equation}
in this equation, $\rho(r)$ denotes the total mass density profile, encompassing both luminous and dark matter components, while $\nu(r)$ represents the density distribution of luminous matter. The anisotropy parameter $\beta(r)$ characterizes the orbital structure of the velocity dispersion, with $\sigma_\theta$ and $\sigma_r$ corresponding to the tangential and radial components, respectively. It is important to note that the power-law index $\gamma$ in the mass density profile is distinct from the PPN parameter $\gamma_{\rm PPN}$. Throughout this work, $\gamma$ (without subscript) always denotes the logarithmic slope of the total mass density distribution, unless otherwise stated. In the special case where $\gamma = \delta = 2$ and $\beta = 0$, this model reduces to the Singular Isothermal Sphere (SIS) profile, a widely adopted approximation in SGL studies. 

In-depth studies of nearby elliptical galaxies have shown that the anisotropy parameter $\beta$ follows a Gaussian distribution, with $\beta = 0.18 \pm 0.13$. This value has been widely adopted in subsequent studies. Accordingly, we marginalize over $\beta$ using a Gaussian prior of $\beta = 0.18 \pm 0.13$ over the range $[\bar{\beta} - 2\sigma_{\beta}, \bar{\beta} + 2\sigma_{\beta}]$, where $\bar{\beta} = 0.18$ and $\sigma_{\beta} = 0.13$ \cite{1995MNRAS.276.1341J,2019MNRAS.488.3745C}.
It should be noted that the parameter $\gamma$ has been reported in previous studies to be redshift-dependent, namely:
\begin{eqnarray}
  \quad P_1: \gamma &=& \gamma_0, \\ \nonumber
\quad P_2: \gamma  & =& \gamma_0 + \gamma_z z_l, 
\end{eqnarray}
here, $\gamma_0$ and $\gamma_z$ are free constants. The $P_1$ parameterization assumes no dependence on redshift or mass distribution and has been widely adopted in previous studies \cite{2019MNRAS.488.3745C}. The $P_2$ model introduces a redshift dependence, allowing the total mass density profile of lens galaxies to evolve with redshift, where $z_l$ denotes the lens redshift.

By solving the Jeans equation within the framework of the lens mass model described in Eq. (\ref{eq:lens_mass_model}), we obtain the following expression for the dynamical mass:
\begin{equation} \label{eq:Mdyn}
M^E_{\rm{dyn}} = \frac{\sqrt{\pi}}{2G}\sigma^2_{\parallel}(\leq R_A) R_E f^{-1}( \gamma, \delta, \beta )
\left( \frac{R_A}{R_E} \right)^{\gamma-2},
\end{equation}
where $\sigma^2_{\parallel}(\leq R_A)$ represents the luminosity-weighted line-of-sight velocity dispersion of the lens galaxy, averaged over the spectroscopic aperture of radius $R_A$. The physical aperture radius and Einstein radius are given by $R_A = D_l \theta_A$ and $R_E = D_l \theta_E$, respectively.
Defining $\xi \equiv \gamma + \delta - 2$, the dimensionless coefficient $f(\gamma,\delta,\beta)$ is given by:
\begin{equation}
\begin{aligned}
f(\gamma ,\delta ,\beta)=&\frac{3-\delta }{(\xi -2\beta )(3-\xi )}\left[ \frac{\Gamma (\frac{\xi -1}{2})}{\Gamma (\frac{\xi }{2})}-\beta \frac{\Gamma (\frac{\xi +1}{2})}{\Gamma (\frac{\xi +2}{2})} \right]\\&\times \frac{\Gamma (\frac{\gamma }{2})\Gamma (\frac{\delta }{2})}{\Gamma (\frac{\gamma -1}{2})\Gamma (\frac{\delta -1}{2})},
\end{aligned}
\end{equation}
where $\Gamma$ denotes the Gamma function.

From the combination of Eqs.~\eqref{eq:Mgrl_Mdyn}, ~\eqref{eq:3}, and ~\eqref{eq:Mdyn}, we obtain the following expression for the velocity dispersion:
\begin{equation} \label{eq:VD_th_RA}
\sigma^2_{\parallel}(\leq R_A) = \frac{c^2}{2\sqrt{\pi}} \frac{2}{1+\gamma_{\text{PPN}}} \frac{D_s}{D_{ls}} \theta_E  f( \gamma, \delta, \beta ) \left( \frac{\theta_A}{\theta_E} \right)^{2 - \gamma}.
\end{equation}

To enable a direct comparison, both the measured velocity dispersion and the model predictions are normalized to a consistent physical aperture, typically taken as $\theta_{\mathrm{eff}}/2$ (where $\theta_{\mathrm{eff}}$ denotes the half-light radius of the lens galaxy). This is achieved by applying a correction to the luminosity-weighted line-of-sight velocity dispersion $\sigma_{\mathrm{ap}}$ obtained from an aperture of size $\theta_{\mathrm{ap}}$, using the following relation:
\begin{equation}\label{eq:VD_obs}
\sigma_{\parallel}^{\rm{obs}} \equiv \sigma_{{\rm{e2}}} = \sigma_{\rm{ap}}[\theta_{\rm{eff}}/(2\theta_{\rm{ap}})]^{\eta},
\end{equation}
Based on the effective radius $R_{\mathrm{eff}}$ of the lens galaxy and its distance $D_l$, the effective angular radius is defined as $\theta_{\mathrm{eff}} = R_{\mathrm{eff}} / D_l$. For the aperture correction exponent of velocity dispersion, we adopt $\eta = -0.066 \pm 0.035$ as determined by \cite{2006MNRAS.366.1126C}.  With this, the total uncertainty of the observed velocity dispersion $\sigma_{\mathrm{obs}}^0$ can be calculated using the following expression:
\begin{equation} \label{eq:VD_th}
\sigma^{\rm{th}}_{\parallel}= \left[\frac{c^2}{2\sqrt{\pi}} \frac{2}{1+\gamma_{\text{PPN}}} \frac{D_s}{D_{ls}} \theta_E f( \gamma, \delta, \beta ) \left( \frac{\theta_{\rm{eff}}}{2\theta_E} \right)^{2 - \gamma}\right]^{1/2}.
\end{equation}

Considering the additional mass contribution from structures along the line of sight, we adopt a $3\%$ fractional uncertainty in the velocity dispersion as the systematic error, following \cite{2007ApJ...671.1568J}. The total uncertainty of the observed $ \sigma^{\rm{th}}_{\parallel} $ is then obtained by combining the statistical and systematic errors in quadrature:
\begin{equation}\label{eq:error}
\begin{aligned}
(\Delta \sigma_0)^2 =& (\Delta \sigma_0^{\text{stat}})^2 + (\Delta \sigma_0^{\text{AC}})^2 + (\Delta \sigma_0^{\text{sys}})^2 \\=& \left[ \frac{\Delta \sigma_{\text{ap}}^2}{\sigma_{\text{ap}}^2} + 0.03^2 + \left( \ln \left( \frac{\theta_{\text{eff}}}{2\theta_{\text{ap}}} \right) \Delta \eta \right)^2 \right] \sigma_0^2,
\end{aligned}
\end{equation}
equation presents the total uncertainty of the observed velocity dispersion $\sigma_0$, which is composed of three independent components: statistical uncertainty $(\Delta \sigma_0^{\text{stat}})^2$, aperture correction uncertainty $(\Delta \sigma_0^{\text{AC}})^2$, and systematic uncertainty $(\Delta \sigma_0^{\text{sys}})^2$. These components are combined in quadrature to yield the total variance. The statistical term arises from the measurement error in $\sigma_{\text{ap}}$, given by $\Delta \sigma_{\text{ap}}^2 / \sigma_{\text{ap}}^2$. The aperture correction uncertainty accounts for the error in the correction exponent $\eta$, scaled by the logarithmic ratio of the effective aperture size $\theta_{\mathrm{eff}}$ to the measurement aperture $\theta_{\mathrm{ap}}$.

With Equation (\ref{eq:VD_th}) and the observed velocity dispersions, we can now constrain $\gamma_{\text{PPN}}$ using existing SGL data. The SGL sample used in this work is presented below.

\subsection{SGL Data Sample}

In this work, we utilize the currently galaxy-scale SGL sample compiled by \cite{2019MNRAS.488.3745C}, which comprises 161 early-type lens galaxies with both high-resolution imaging and stellar velocity dispersion measurements. This sample integrates lens systems from six major surveys: the Lens Structure and Dynamics (LSD) survey \cite{2002ApJ...568L...5K,2003ApJ...599...70K,2002ApJ...575...87T,2004ApJ...611..739T}, the Strong Lensing Legacy Survey \cite{2011ApJ...727...96R,2013ApJ...777...98S,2015ApJ...800...94S}, the Sloan Lens ACS (SLACS) survey \cite{2009ApJ...705.1099A,2010ApJ...724..511A,2015ApJ...803...71S,2017ApJ...851...48S}, the SLACS extension known as ``SLACS for the Masses" , the Baryon Oscillation Spectroscopic Survey (BOSS) Emission-Line Lens Survey \cite{2012ApJ...744...41B,2016ApJ...824...86S,2016ApJ...833..264S}.

The sample selection adheres to rigorous criteria to ensure the validity of the spherical symmetry assumption for lens galaxies: (i) all lens galaxies are early-type galaxies with E/S0 morphologies; (ii) they exhibit no significant substructure or close massive companions. The full sample covers lens redshifts in the range of approximately $0.06 < z_l < 1.0$ and source redshifts spanning $0.2 < z_s < 3.5$, with velocity dispersions predominantly distributed within $180 \,\mathrm{km/s} < \sigma_{\mathrm{e2}} < 300 \,\mathrm{km/s}$. For each system, the observational data include the lens redshift $z_l$, source redshift $z_s$, Einstein angle $\theta_E$, effective (half-light) angular radius $\theta_{\mathrm{eff}}$, the velocity dispersion measured within an aperture $\sigma_{\mathrm{ap}}$, and the corresponding aperture angular radius $\theta_{\mathrm{ap}}$. Detailed information for all 161 systems is provided in the Appendix of \cite{2019MNRAS.488.3745C}.

Given that the BAO data employed in this work have a maximum redshift of $z=2.334$, we applied a selection cut to the original sample to ensure consistency in redshift coverage between the gravitational lensing sample and the BAO data. Specifically, we excluded lens systems with source redshift $z_s > 2.334$, as systems beyond the BAO redshift range cannot provide effective cosmological constraints in our joint analysis. Following this selection, we obtained 120 SGL systems satisfying the redshift consistency requirement. The redshift distribution of the selected sample remains consistent with that of the original sample, with lens redshifts still predominantly distributed in the range $0.06 < z_l < 0.73$, and the velocity dispersion distribution exhibits no significant alteration, thereby preserving the statistical integrity of the sample. These 120 systems constitute the foundation of our joint analysis in this work.

Since SGL observations only provide the lens and source redshifts, they do not directly yield the angular diameter distances required in the analysis. In conventional approaches, these distances are typically inferred by assuming a fiducial cosmological model. However, this inevitably introduces cosmological model dependence. To avoid this limitation, we instead employ BAO measurements to reconstruct the distance--redshift relation in a model-independent manner.

\subsection{BAO angular scale measurements}

As discussed above, the interpretation of SGL observations relies on the knowledge of angular diameter distances, which cannot be directly obtained from lensing data alone. To address this issue without assuming a fiducial cosmological model, we employ BAO observations as a standard ruler to reconstruct the distance--redshift relation.

BAO imprint a characteristic scale in the large-scale structure of the Universe, serving as a natural standard ruler for cosmology. This intrinsic scale is set by the comoving sound horizon at the drag epoch, $r_d \approx 150~\mathrm{Mpc}$, which is frozen when baryons decouple from photons at redshift $z_d$. The sound horizon is given by
\begin{equation}
r_d = \int_{z_d}^{\infty} \frac{c_s(z)}{H(z)} \, dz, \label{eq:r_d}
\end{equation}
where $c_s(z)$ is the sound speed of the photon-baryon fluid and $H(z)$ is the Hubble parameter. The observed angular BAO scale $\theta_{\rm BAO}$ is related to the angular diameter distance $D^A$ by
\begin{equation}
\theta_{\rm BAO} = \frac{r_d}{(1+z) D^A}. \label{eq:theta_BAO}
\end{equation}
This relation allows angular diameter distances to be inferred from BAO measurements in a model-independent way, while still requiring the sound horizon scale \(r_d\) as an input.

In this work, we utilize three complementary BAO datasets. First, we adopt 15 measurements of the transverse BAO angular scale (2D-BAO) from \cite{2020MNRAS.497.2133N}, derived from SDSS DR7, DR10, DR11, DR12, and DR12Q. These measurements are obtained without assuming a fiducial cosmological model, thus providing a cosmology-independent distance anchor. Second, we include anisotropic BAO measurements (3D-BAO) from DES Year 6 combined with BOSS/eBOSS \cite{2017MNRAS.470.2617A,2021PhRvD.103h3533A}. Third, we incorporate the latest BAO results from the DESI DR2 \cite{2025PhRvD.112h3515A}, which represent a significant advancement over previous datasets.

\begin{figure}
{\includegraphics[width=1\linewidth]{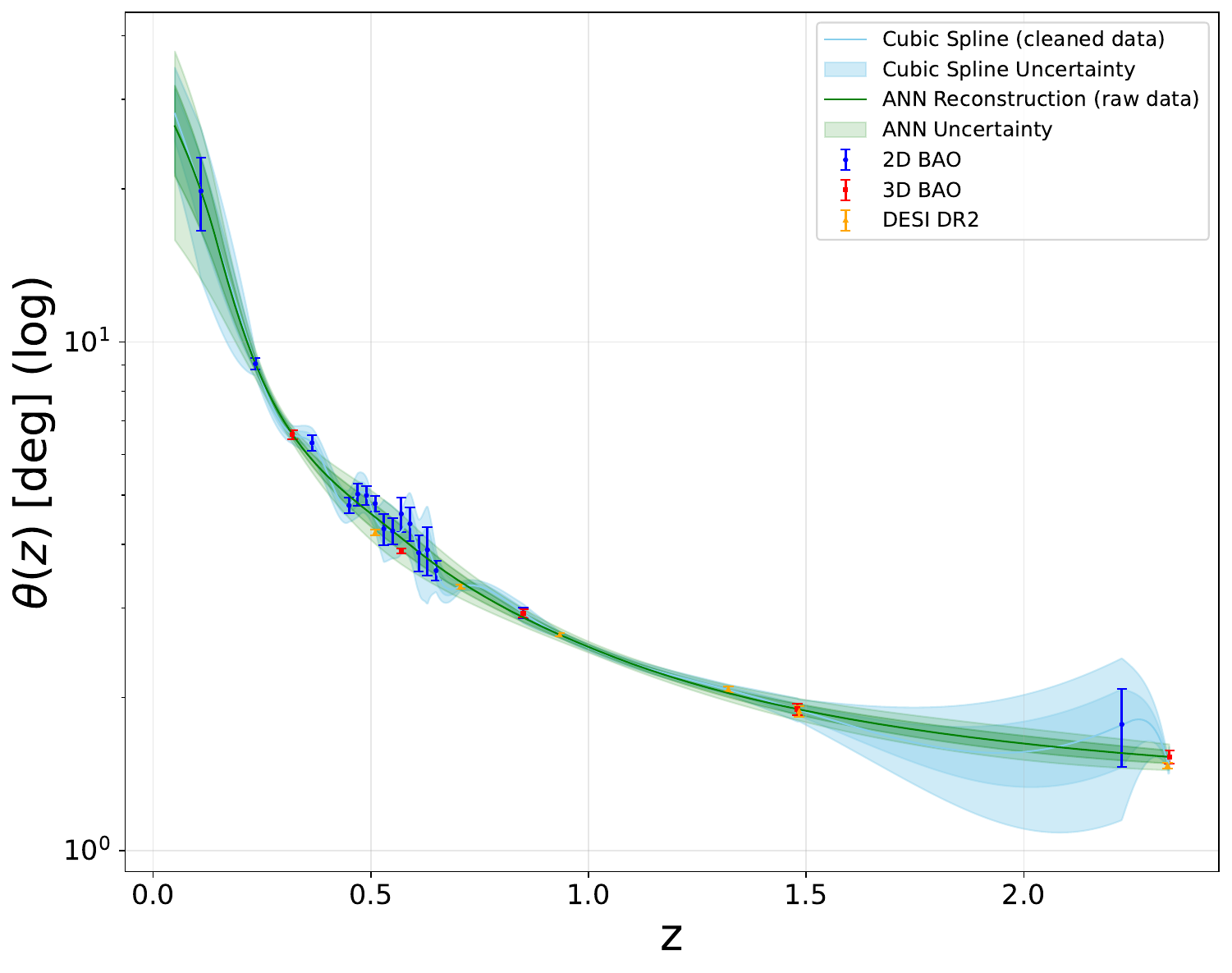}}
\caption{Comparison of BAO angular scale reconstructions using different methods: cubic spline with blue line, ANN with green line, alongside 2D BAO, 3D BAO, and DESI DR2 measurements with error bars. The shaded regions represent the corresponding $1\sigma$ prediction uncertainties.}\label{fig1}
\end{figure}

The DESI DR2 measurements offer substantially improved precision, with the sample size doubled compared to DESI R1 releases \cite{2025JCAP...02..021A}, leading to an overall precision increase of $30\%$--$50\%$ \cite{2025PhRvD.112h3515A}. Key improvements include sub-percent precision in the nearby universe, an unprecedented $0.46\%$ accuracy at intermediate redshifts, and the first two-dimensional anisotropic BAO constraints from quasars. Collectively, these data provide a distance ladder spanning nine billion years of cosmic evolution with exceptional precision, establishing a robust foundation for testing cosmological models. The complete set of BAO angular scale measurements (2D-BAO and 3D-BAO) used in this work is shown in Fig.~\ref{fig1}.

\textit{ANN Reconstruction:} 
To reconstruct the redshift evolution of the BAO angular scale $\theta(z)$ without assuming a specific functional form, we employ an artificial neural network (ANN) as a non-parametric reconstruction method. Our implementation is based on the publicly available ReFANN code \cite{2020ApJS..246...13W,2025ApJ...987...58W}, which has been widely applied in cosmological analyses.

The observational data are organized as $(z_i, \theta_i, \sigma_i)$, where $z_i$ denotes the redshift, $\theta_i$ the measured BAO angular scale, and $\sigma_i$ the corresponding uncertainty. The network takes $z$ as input and outputs the reconstructed function $\theta(z)$, with parameters optimized by minimizing a loss function weighted by observational errors.

The adopted network architecture is sufficiently flexible to capture the underlying functional dependence of $\theta(z)$, while avoiding overfitting. After training, the reconstructed function is evaluated on a dense redshift grid to obtain a smooth representation of $\theta(z)$, together with its associated uncertainty. The ANN reconstruction is shown in Fig.~\ref{fig1}, where the green curve represents the reconstruced result and the shaded region denotes the corresponding confidence interval. These results indicate that the ANN provides an accurate description of the BAO angular scale data without significant overfitting.
The use of this non-parametric approach allows us to reconstruct $\theta(z)$ in a model-independent manner. In combination with the cubic spline method described below, this enables a robust assessment of potential reconstruction-related systematics.

\textit{Cubic Spline Reconstruction:} Following the approach of Bernal, Verde, and Riess \cite{2016JCAP...10..019B} and Aylor et al. \cite{2019ApJ...874....4A}, the second reconstruction method adopted in this work is based on CS reconstruction. This technique reconstructs the redshift evolution of the BAO angular scale $\theta(z)$ using piecewise cubic polynomials, which are continuous in the function as well as in their first and second derivatives. As a result, the reconstructed curve is smooth while avoiding overfitting to observational noise.

In practice, the observational data $(z_i, \theta_i, \sigma_{\theta,i})$ are first sorted by redshift to ensure strictly increasing redshift values, which is required for the spline reconstruction. The CS is then fitted to the data by minimizing a weighted $\chi^2$. To estimate the uncertainty of the reconstructed function, the observational errors are mapped onto the reconstruction grid using linear reconstruction. The resulting CS reconstruction thus captures the overall trend of the BAO angular scale while providing a corresponding uncertainty estimate. Figure~\ref{fig1} shows the CS based on the cleaned data (blue curve with shaded region) in comparison with the ANN reconstruction based on the raw data (green curve with shaded region). 

\subsection{The Likelihood Function}

We employ the \texttt{emcee} Python module developed by Foreman-Mackey \cite{2013PASP..125..306F}, which implements a Markov Chain Monte Carlo (MCMC) sampler, to maximize the likelihood $L \propto e^{-\chi^{2}/2}$. The PPN parameter $\gamma_{\mathrm{PPN}}$ and the lens model parameters $(\gamma_{0}, \gamma_{z})$ are fitted simultaneously. The corresponding $\chi^{2}$ function is defined as:

\begin{equation}
\chi^2 (\mathbf{p}, \gamma_{\text{PPN}}) = \sum_{i=1}^{120} \frac{\left( \sigma_0^{\text{th}} (z_i; \gamma_{\text{PPN}}) - \sigma_0^{\text{obs}} (z_i; \mathbf{p}) \right)^2}{\Delta \sigma_0^{\text{tot}} (z_i)^2},
\label{eq:chi2}
\end{equation}
where $\mathbf{p}$ denotes the parameters of the lens model. The total uncertainty $\Delta \sigma_{0}^{\mathrm{tot}}$ is obtained by combining the observational uncertainty $\Delta \sigma_{0}^{\text{SGL}}$, given by Eq.~(\ref{eq:error}), with the uncertainty from the BAO distance calibration, $\Delta \sigma_{0}^{\text{BAO}}$. Thus:
\begin{equation}
\left( \Delta \sigma_0^{\text{tot}} \right)^2 = \left( \Delta \sigma_0^{\text{SGL}} \right)^2 + \left( \Delta \sigma_0^{\text{BAO}} \right)^2.
\label{eq:error_propagation}
\end{equation}

From Eq.~(\ref{eq:VD_th}), the contribution of the velocity dispersion uncertainty to the gravitational lensing distance ratio can be written as:
\begin{equation}
\Delta \sigma_0^{\text{BAO}} = \sigma_0^{\text{th}} \frac{\Delta D_{\text{ratio}}}{2D_{\text{ratio}}}
\label{eq:BAO_error},
\end{equation}
here $D_{\mathrm{ratio}} \equiv D_{s}/D_{ls}$ denotes the distance ratio defined in Eq.~(\ref{eq:VD_th}), with its uncertainty represented by $\Delta D_{\mathrm{ratio}}$. By reconstructing a smooth angular–redshift relation from BAO data, the angular diameter distances ratio  $Ds/D_{ls}$ for each SGL system, can be determined by the standard distance relation between the lens and the source for a spatially flat universe \cite{1972gcpa.book.....W}:
\begin{equation}
D_{\text{ratio}} \label{eq:distance_ratio}
= \frac{D_s (1+z_s)}{D_s (1+z_s) - D_l (1+z_l)},
\end{equation}
by combining Eqs.~(\ref{eq:theta_BAO}) and ~(\ref{eq:distance_ratio}), the relation between $\theta_{\mathrm{BAO}}$ and $D_{\mathrm{ratio}}$ can be obtained as:
\begin{equation}
D_{\text{ratio}} =  \frac{\theta_l}{\theta_l - \theta_s},
\label{eq:BAO_ratio}
\end{equation}
where $\theta_l \equiv \theta_{\rm BAO}(z_l)$ and $\theta_s \equiv \theta_{\rm BAO}(z_s)$ denote the BAO angular scales evaluated at the lens and source redshifts, respectively. The corresponding uncertainty can be written as:
\begin{equation}
\Delta D_{\text{ratio}}  = \frac{D_{\text{ratio}}^2}{\theta_l} \sqrt{ \Delta\theta_s^2 + \left( \frac{\theta_s}{\theta_l} \right)^2 \Delta\theta_l^2 }.
\end{equation}
From the above equation, it can be seen that our tests of GR are independent of the cosmological model, the Hubble constant, the sound horizon, and the equation of state of dark energy. This can significantly reduce the systematic bias introduced by cosmological priors in tests of GR.

\section{Results and Discussion}
Based on the likelihood function constructed from Eq.~(\ref{eq:chi2}), we perform a joint parameter estimation using the MCMC method implemented in emcee. We simultaneously constrain the PPN parameter $\gamma_{\rm PPN}$ and the parameters describing the mass-density distribution of the lens galaxies. In this section, we present the results obtained under two different mass-density parameterizations, denoted as the $P_1$ and $P_2$ models, respectively. We also compare the differences between the two BAO reconstruction approaches adopted in this work, namely the ANN and the CS reconstruction.

\subsection{Results for the \texorpdfstring{$P_{1}$}{P1} Model}
\begin{figure}
{\includegraphics[width=1\linewidth]{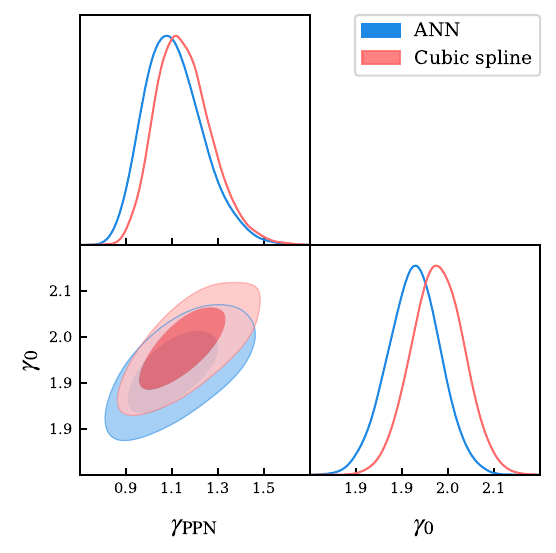}}
\caption{Posterior distributions of $(\gamma_{\rm PPN},\gamma_0)$ for the $P_1$ model. The contours show the 68\% and 95\% confidence regions. Blue contours correspond to the ANN reconstruction using raw data, while red contours correspond to the cubic spline reconstruction. }\label{fig2}
\end{figure}

We first consider the $P_1$ parameterization, in which the lens galaxy mass-density slope is assumed to be constant, $\gamma = \gamma_0$. 
Figure~\ref{fig2} presents the posterior distributions of $(\gamma_{\rm PPN}, \gamma_0)$.

Using the ANN-based BAO reconstruction, we obtain $\gamma_{\rm PPN} = 1.102^{+0.148}_{-0.125}$, $\gamma_0 = 1.966^{+0.033}_{-0.035}$. For the CS, the results are $\gamma_{\rm PPN} = 1.150^{+0.139}_{-0.118}$, $\gamma_0 = 1.995^{+0.034}_{-0.034}$. The constraints obtained from the two BAO reconstruction methods exhibit a high degree of overlap within the $1\sigma$ confidence intervals. Although different reconstruction techniques are adopted, the two approaches yield comparable levels of precision, indicating that the choice of distance reconstruction method does not have a significant impact on the final results. For the $P_1$ model, the best-fit value of $\gamma_0$ is close to $2$, corresponding to an approximately isothermal density profile. This is consistent with previous observational studies based on SGL systems. Regarding the PPN parameter, the results obtained using the ANN reconstruction are consistent with the prediction of GR, $\gamma_{\rm PPN}=1$, at the $1\sigma$ confidence level. The results derived from the CS are also consistent with the GR prediction within $2\sigma$. 
These results indicate that, under the $P_1$ assumption, the current observational data do not provide evidence for deviations from GR on galactic scales.

In comparison with the results of \cite{2022ApJ...927...28L}, although our constraints on $\gamma_{\rm PPN}$ are slightly less precise, this difference can be attributed to the size of the BAO reconstruction dataset. In this work, the BAO reconstruction is based on 24 data points, whereas \citet{2022ApJ...927...28L} utilized a significantly larger sample of 1048 Type Ia supernovae. Therefore, despite the smaller dataset, our analysis demonstrates that a model-independent BAO reconstruction combined with SGL observations can still provide robust constraints on gravity.

\subsection{Results for the \texorpdfstring{$P_{2}$}{P2} Model}

\begin{figure}
{\includegraphics[width=1\linewidth]{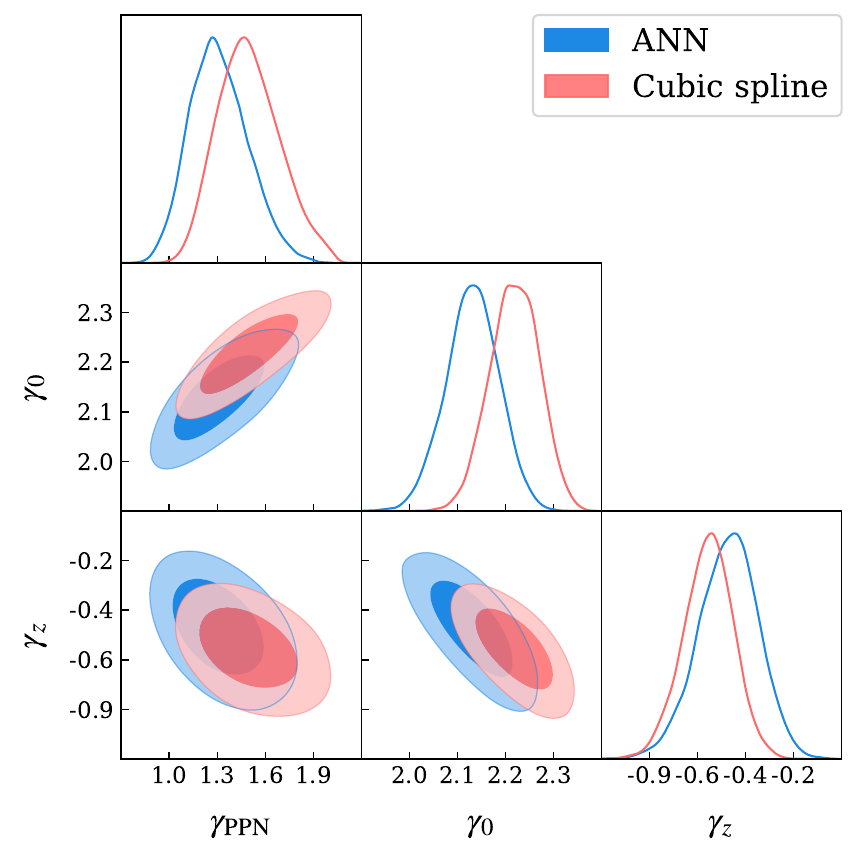}}
\caption{Posterior distributions of $(\gamma_{\rm PPN},\gamma_0,\gamma_z)$ for the $P_2$ model. The contours show the 68\% and 95\% confidence regions. Blue contours correspond to the ANN reconstruction using raw data, while red contours correspond to the cubic spline reconstruction. }\label{fig3}
\end{figure}

We next consider the $P_2$ parameterization, in which the mass-density slope of the lens galaxies is allowed to evolve with redshift. The parameterization is given by $\gamma = \gamma_0 + \gamma_z z_l$. Figure~\ref{fig3} presents the joint posterior distributions of $(\gamma_{\rm PPN}, \gamma_0, \gamma_z)$. Using the ANN reconstruction method, we obtain $\gamma_{\rm PPN} = 1.315^{+0.181}_{-0.155}$, $\gamma_0 = 2.145^{+0.050}_{-0.050}$, $\gamma_z = -0.515^{+0.119}_{-0.130}$.

For the CS method, the corresponding results are $\gamma_{\rm PPN} = 1.485^{+0.193}_{-0.168}$, $\gamma_0 = 2.225^{+0.046}_{-0.048}$, $\gamma_z = -0.601^{+0.100}_{-0.109}$. Compared with the $P_1$ model, introducing the redshift-dependent parameter $\gamma_z$ slightly increases the best-fit value of $\gamma_0$. Meanwhile, the negative value of $\gamma_z$ suggests a mild evolution of the mass-density slope with redshift, implying that the density profile may become somewhat shallower at higher redshift. This behavior is qualitatively consistent with previous studies indicating possible evolution in the internal structure of early-type galaxies.

For the PPN parameter, the $P_2$ model yields larger best-fit values of $\gamma_{\rm PPN}$ compared with the $P_1$ case. The ANN constraint remains compatible with GR at the $2\sigma$ level. The CS constraint deviates from GR at approximately the $2.5\sigma$ level, indicating mild tension. Given the current uncertainties, the present dataset does not provide strong statistical evidence for deviations from GR even when allowing for redshift evolution in the mass-density profile.

\begin{table*}
 \renewcommand{\arraystretch}{2.0}
    \setlength{\tabcolsep}{10pt}
\caption{Parameter constraints obtained from different BAO reconstruction methods under the $P_1$ and $P_2$ models.}
\label{table:results}
\centering
\scalebox{1}{
\begin{tabular}{c|c|c|c|c|c}
\hline
\hline
Model & Reconstruction Method & $\gamma_{\rm PPN}$ & $\gamma_0$ & $\gamma_z$ &  BIC  \\
\hline

\multirow{2}{*}{$P_1$}
& ANN & $1.102^{+0.148}_{-0.125}$ & $1.966^{+0.033}_{-0.035}$ & --- & 347.092 \\
& Cubic spline & $1.150^{+0.139}_{-0.118}$ & $1.995^{+0.034}_{-0.034}$ & --- & 349.726 \\

\hline

\multirow{2}{*}{$P_2$}
& ANN & $1.315^{+0.181}_{-0.155}$ & $2.145^{+0.050}_{-0.050}$ & $-0.515^{+0.119}_{-0.130}$ & 376.889 \\
& Cubic spline & $1.485^{+0.193}_{-0.168}$ & $2.225^{+0.046}_{-0.048}$ & $-0.601^{+0.100}_{-0.109}$ & 365.518 \\

\hline
\hline
\end{tabular}
}
\end{table*}

\subsection{Comparison Between Reconstruction Methods}

It is also instructive to compare the results obtained from the two BAO reconstruction techniques. As shown in Table~\ref{table:results}, both the ANN and CS produce consistent parameter constraints for both the $P_1$ and $P_2$ models. Although the ANN reconstruction uses the raw BAO data while the CS method is applied to the cleaned dataset, the resulting posterior distributions show significant overlap. This agreement indicates that the final constraints on $\gamma_{\rm PPN}$ are relatively insensitive to the specific reconstruction method used to infer the BAO angular scale. In other words, the dominant uncertainties in the analysis arise primarily from the observational errors of the lensing systems rather than from the BAO reconstruction procedure.

We further assess the relative performance of the two reconstruction methods using the BIC values listed in Table~\ref{table:results}. The cubic spline reconstruction yields slightly lower BIC values than the ANN reconstruction for both lens mass models: $\Delta\mathrm{BIC} = 2.6$ under the $P_1$ model and $\Delta\mathrm{BIC} = 11.4$ under the $P_2$ model. While the difference under $P_1$ is negligible, the more pronounced difference under $P_2$ suggests that the cubic spline reconstruction provides a marginally better description of the BAO data when combined with the more complex lens model. Nevertheless, given the overall consistency of the $\gamma_{\rm PPN}$ constraints between the two reconstruction methods, the choice of BAO reconstruction technique does not significantly impact the conclusions of this work.

Based on the same BIC comparison, the $P_1$ model is strongly favored over the $P_2$ model under both reconstruction techniques, with $\Delta\mathrm{BIC} = 29.8$ for the ANN reconstruction and $\Delta\mathrm{BIC} = 15.8$ for the cubic spline reconstruction. These differences indicate that the constant-density-slope assumption is statistically preferred for the current sample, and introducing redshift evolution in the density profile is not justified by the data.

\subsection{Comparison of Different Survey Samples and Consistency Analysis}

\begin{figure*}
\centering
\begin{minipage}{0.46\textwidth}
\centering
\includegraphics[width=\linewidth]{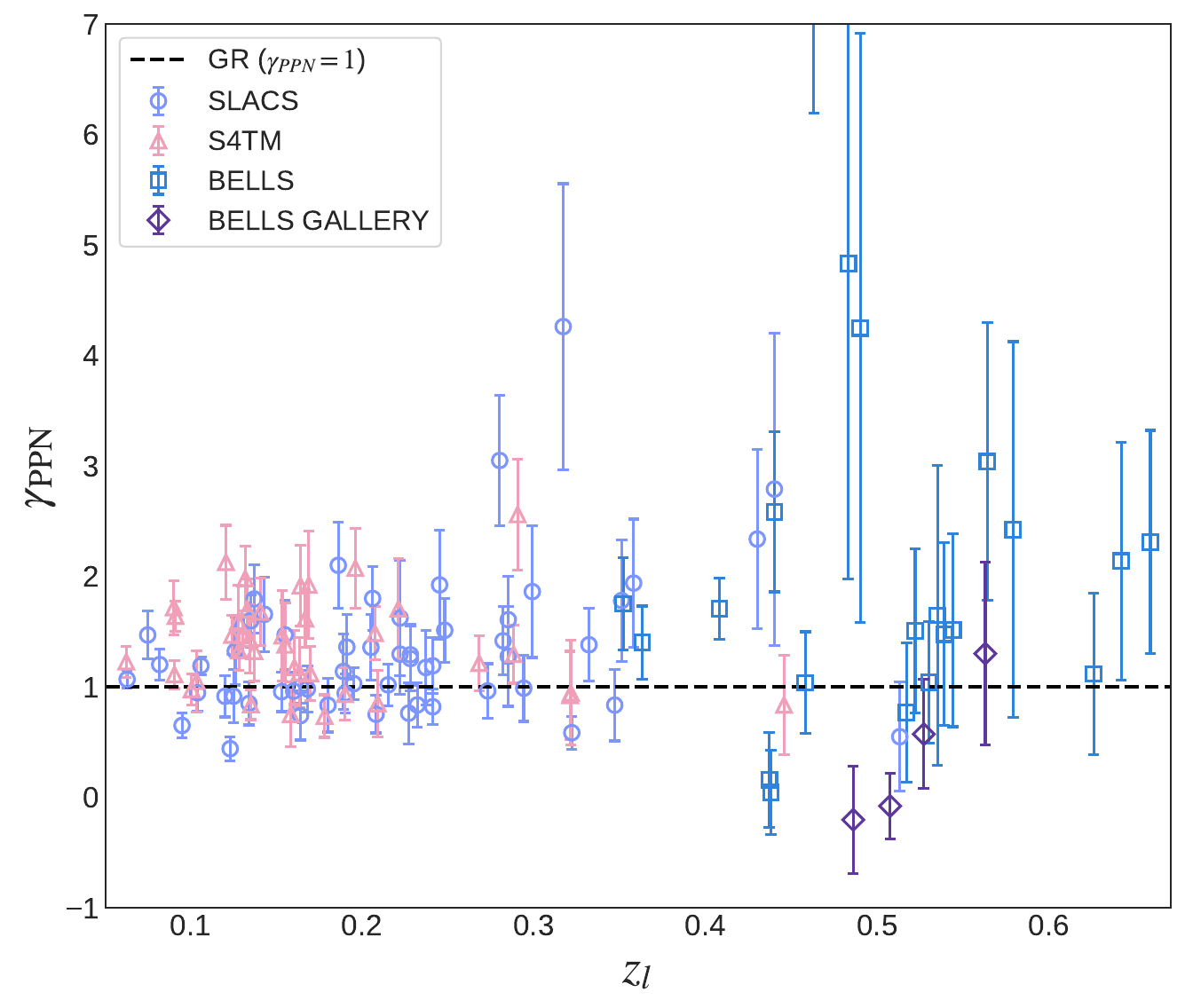}
\caption{Based on the P1 model ($\gamma=2$), the distribution of $\gamma_{\rm PPN}$ estimates inferred from 120 SGL systems as a function of lens redshift $z_l$. The horizontal dashed line indicates the GR prediction ($\gamma_{\rm PPN}=1$). Data points from different surveys are distinguished by different markers.}
\label{fig:P4}
\end{minipage}
\hfill
\begin{minipage}{0.46\textwidth}
\centering
\raisebox{-1cm}{ 
\includegraphics[width=\linewidth]{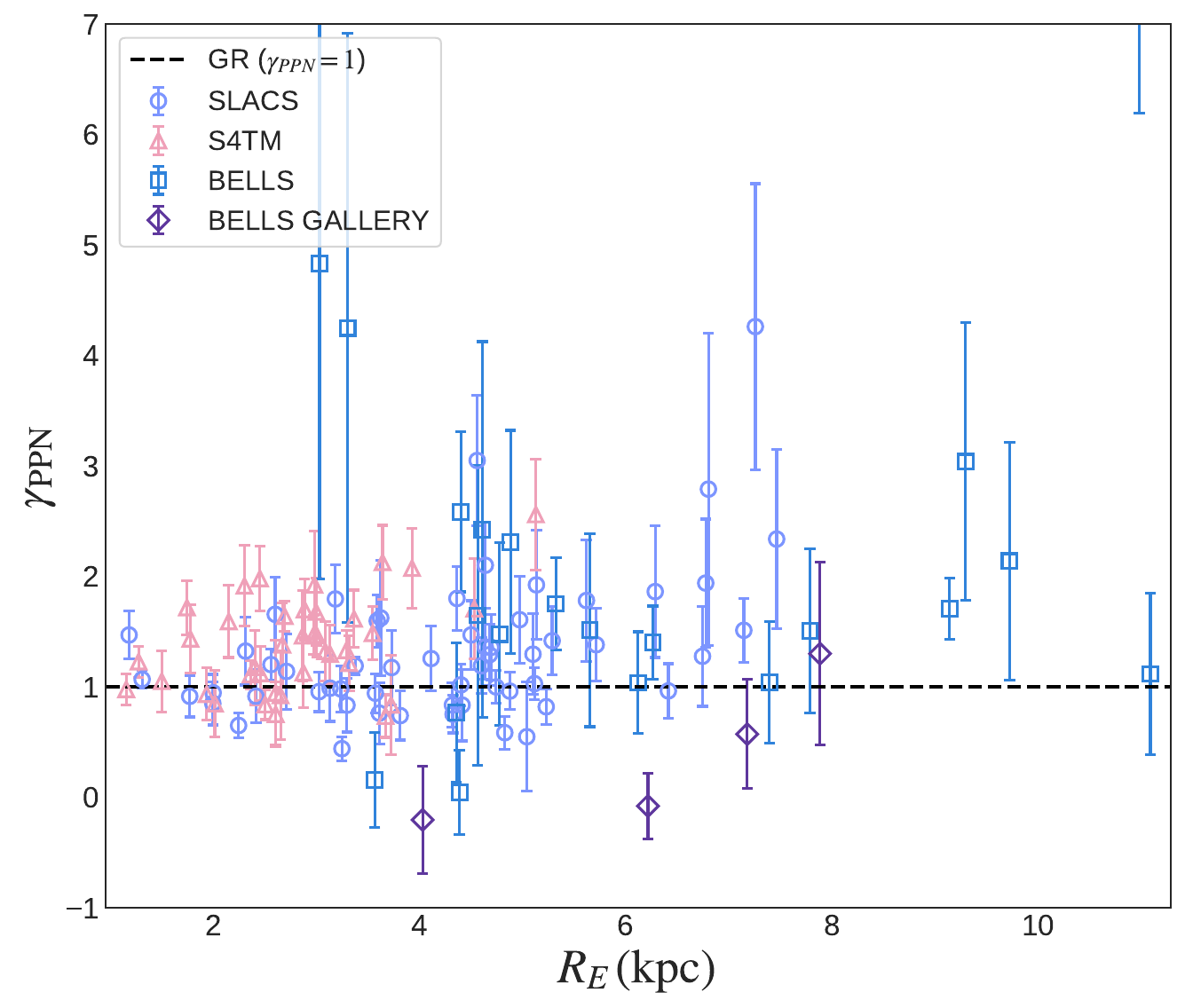}
}
\caption{Based on the P1 model ($\gamma=2$), the distribution of $\gamma_{\rm PPN}$ estimates inferred from 120 SGL systems as a function of Einstein radius $R_E$. The horizontal dashed line indicates the GR prediction ($\gamma_{\rm PPN}=1$). Data points from different surveys are distinguished by different markers.}
\label{fig4}
\end{minipage}

\end{figure*}

Figure~\ref{fig:P4} presents the distribution of the individually fitted $\gamma_{\rm PPN}$ as a function of lens redshift $z_l$ for the 120 SGL systems. For the majority of the sample, the best-fit values of $\gamma_{\rm PPN}$ and their corresponding $1\sigma$ uncertainties are consistent with the general relativistic prediction ($\gamma_{\rm PPN}=1$). The data points are symmetrically distributed around $\gamma_{\rm PPN}=1$ and do not exhibit any systematic trend with redshift. This result provides support for the validity of GR on galactic scales.

The lens systems from the SLACS and S4TM surveys show good agreement in their overlapping redshift range ($z_l \approx 0.1 - 0.4$), with relatively small uncertainties and consistent parameter estimates.

In contrast, several systems from the BELLS survey display larger scatter, with a few cases deviating noticeably from $\gamma_{\rm PPN}=1$ and exhibiting comparatively larger uncertainties. Taking SDSS J1352+3216 at $z_l = 0.436$ as an example, this system has a relatively large Einstein radius but a comparatively small stellar velocity dispersion. Such a configuration may introduce tension between the lensing-inferred mass and the dynamical mass, thereby biasing the fitted $\gamma_{\rm PPN}$ value. This apparent discrepancy could also indicate that the system may not be fully described by the simple power-law mass density profile assumed in our analysis.

Figure~\ref{fig4} presents the distribution of the individually fitted post-Newtonian parameter $\gamma_{\rm PPN}$ as a function of the Einstein radius $R_E$ (in kpc) for the 120 SGL systems. No significant systematic trend of $\gamma_{\rm PPN}$ with $R_E$ is observed. A few systems that appear to deviate from $\gamma_{\rm PPN}=1$ warrant further high-resolution observations and refined modeling. If such deviations are confirmed, they could indicate new gravitational physics; otherwise, they may reflect complex internal structures of lens galaxies. Overall, the results support the validity of GR on galactic scales from a few kpc up to nearly $10\,\mathrm{kpc}$, with no strong evidence for a scale-dependent variation of gravity within this range.

\section{Conclusion}
In this work, we have developed a cosmological model-independent framework for testing general relativity by combining reconstructed BAO angular scale measurements with galaxy-scale strong gravitational lensing data. Using two non-parametric reconstruction techniques—ANN and CS reconstruction—we infer the angular diameter distances to lenses and sources directly from BAO observations, without imposing any assumptions on the cosmological model, the Hubble constant, the sound horizon, or the dark energy equation of state. This approach minimizes potential systematic biases that may arise from model-dependent distance priors in tests of gravity. Our analysis incorporates the latest BAO measurements from DESI DR2, which significantly improve the precision and redshift coverage of the reconstructed distance relation, providing a more robust foundation for cosmology-independent gravity tests.

Applying this method to a sample of 120 SGL systems, we perform a joint analysis under two lens mass models: a constant-density-slope model ($P_1$) and a redshift-evolving model ($P_2$). As summarized in Table~\ref{table:results}, the $P_1$ model yields $\gamma_{\rm PPN} = 1.102^{+0.148}_{-0.125}$ from the ANN reconstruction and $\gamma_{\rm PPN} = 1.150^{+0.139}_{-0.118}$ from the CS reconstruction, both consistent with the GR prediction ($\gamma_{\rm PPN}=1$) at the $1\sigma$ and $2\sigma$ confidence levels, respectively. For the $P_2$ model, the ANN reconstruction gives $\gamma_{\rm PPN} = 1.315^{+0.181}_{-0.155}$, compatible with GR at the $2\sigma$ level, while the CS reconstruction yields $\gamma_{\rm PPN} = 1.485^{+0.193}_{-0.168}$, showing mild tension at the $\sim2.5\sigma$ level. The BIC values strongly favor the simpler $P_1$ model over $P_2$, with $\Delta\mathrm{BIC}\approx30$ for ANN and $\Delta\mathrm{BIC}\approx16$ for CS, indicating that redshift evolution in the density slope is not statistically justified by the current data.

A comparison between the two reconstruction methods reveals excellent consistency in the inferred parameters, demonstrating that the constraints on $\gamma_{\rm PPN}$ are robust against the choice of BAO reconstruction technique. Quantitatively, the uncertainties in the final constraints are dominated by the SGL observational errors, with the BAO reconstruction contributing only marginally to the total error budget. The inferred values of $\gamma_{\rm PPN}$ are consistent with unity within the corresponding confidence intervals, and no statistically significant evidence for departures from GR is found within the probed redshift and kiloparsec-scale range. Overall, the current observational data support the validity of GR on galactic scales.

Looking ahead, future surveys such as DESI, Euclid, and the Vera C. Rubin Observatory will significantly improve the precision of BAO and SGL measurements, enabling more stringent tests of gravity. In addition, the forthcoming China Space Station Telescope (CSST) will greatly increase the number of well-resolved SGL systems through its wide-field, high-resolution imaging, further enhancing constraints on $\gamma_{\rm PPN}$. Together, these next-generation surveys will substantially improve the statistical power of future tests of GR.

\section*{Acknowledgments}
This work was supported by National Key R$\&$D Program of China (No. 2024YFC2207400).

\bibliography{prd_refers}

\end{document}